\documentclass[aps,pra,twocolumn,amsmath,amssymb,superscriptaddress,floatfix]{revtex4-1}

\usepackage{graphicx}
\usepackage{multirow}
\usepackage{color}
\usepackage{bm}

\newcommand{\beq}{\begin{eqnarray}}
\newcommand{\eeq}{\end{eqnarray}}

\newcommand{\be}{\begin{equation}}
\newcommand{\ee}{\end{equation}}
 
\newcommand{\volintpi}[2]{\int \frac{\mathrm{d}^{#2}\bm{#1}}{{(2\pi)}^{#2}}}
\newcommand{\kk}{\bm{k}}

\newcommand{\td}{\mathrm{d}}
\newcommand{\p}[2]{\frac{\partial {#1}}{\partial {#2}}}
\newcommand{\pa}[1]{\frac{\partial}{\partial {#1}}}
\newcommand{\df}{\p{f^\pm_0}{\varepsilon}}

\newcommand{\dEene}{\bm{E}\cdot \p{\varepsilon^\pm}{\bm{k}}}
\newcommand{\dE}{\bm{E}\cdot \pa{\bm{k}}}

\begin{document}

\title{
Nonlinear spin current generation in noncentrosymmetric 
spin-orbit coupled systems
}

\author{Keita Hamamoto}
\affiliation{Department of Applied Physics, The University of 
Tokyo, Tokyo, 113-8656, Japan}
\author{Motohiko Ezawa}
\affiliation{Department of Applied Physics, The University of 
Tokyo, Tokyo, 113-8656, Japan}
\author{Kun Woo Kim}
\affiliation{School of Physics, Korea Institute for Advanced Study, Seoul 02455, Korea}
\author{Takahiro Morimoto}
\affiliation{Department of Physics, University of California, Berkeley, CA 94720}
\author{Naoto Nagaosa}
\affiliation{Department of Applied Physics, The University of 
Tokyo, Tokyo, 113-8656, Japan}
\affiliation{RIKEN Center for Emergent Matter Science 
(CEMS), Wako, Saitama, 351-0198, Japan}

\date{\today}

\begin{abstract}

Spin current plays a central role in spintronics. 
In particular, finding more efficient ways to generate spin current has been an important issue and studied actively. For example, representative methods of spin current generation include spin polarized current injections from 
ferromagnetic metals, spin Hall effect, and spin battery.
Here we theoretically propose a new mechanism of spin current generation based on nonlinear phenomena.
By using Boltzmann transport theory, we show that a
simple application of the electric field $\bm{E}$ induces spin current proportional to $\bm{E^2}$ in noncentrosymmetric 
spin-orbit coupled systems.
We demonstrate that the nonlinear spin current of the proposed mechanism is supported in the surface state of three-dimensional 
topological insulators and two-dimensional semiconductors 
with the Rashba and/or Dresselhaus interaction.
In the latter case, the angular dependence of the nonlinear spin current can be manipulated by the direction of the electric field and by
the ratio of the Rashba and Dresselhaus interactions.
We find that the magnitude of the spin current largely exceeds those in the previous methods
for a reasonable magnitude of the electric field.
Furthermore, we show that application of AC electric fields (e.g. terahertz light) leads to the rectifying effect of the spin current where DC spin current is generated.
These findings will pave a new route to manipulate the spin current in noncentrosymmetric crystals.
\end{abstract}

\maketitle

\section{Introduction}
Spins and their flow in solids have attracted recent intensive attentions
from the viewpoints of both fundamental physics and spintronics
applications. 
The conventional and direct way to generate spins or spin 
current in solids is to inject the spin polarized current from 
metallic ferromagnets \cite{Datta1990a,Gardelis1999,Schmidt1999,Hu2001}.
Meanwhile, recent researches have been focusing on the electric manipulation of spin and spin current without using the magnets, where the relativistic spin-orbit interaction (SOI) 
plays an essential role.
For such an example, the spin Hall effect supports
the conversion of the charge current to the spin current 
\cite{Hirsch1999,Zhang2000,Murakami2003,Kato2004,
Murakami2004,Murakami2004a,Sinova2004,Wunderlich2005,Engel2005,
Sugimoto2006,Valenzuela2006,Sinova2015}. 
In the presence of the SOI, the spin Hall conductivity $\sigma_H^s$ becomes nonzero
due to the extrinsic mechanism such as the skew scattering 
\cite{Hirsch1999,Zhang2000,Kato2004} or the
intrinsic mechanism by the Berry phase of the Bloch wave functions 
\cite{Murakami2003,Sinova2004,Murakami2004,Murakami2004a,Wunderlich2005}.
These two mechanisms induce the $\sigma_H^s$ proportional to 
$O(\tau)$ and $O(1)$, respectively, in terms of the transport lifetime $\tau$. 
Spin battery is another method to produce the spin current, where the
precession of the ferromagnetic moment is excited by the 
magnetic resonance absorption, and the damping of this 
collective mode results in the flow of the spin current 
to the neighboring system through the interface \cite{Saitoh2006a,Ando2011,Dushenko2016,Lesne2016,Kondou2016}.
Interband spin selective optical transition under the irradiation of the circularly polarized light also induces the spin polarized current which is known as the circular photogalvanic effect \cite{Ganichev2001,Ganichev2002a}. 
These methods have been successfully applied to study the variety of phenomena, 
but the experimental signals associated with the spin current are quite small and the device structure to detect them is limited.
A more efficient way to create the spin current based on another physical origin
has been desired for the purpose of spintronics application.

In this paper, we theoretically propose that a simple
application of the electric field produces the nonlinear spin current
proportional to the square of the electric field ($E^2$) and also the square of the transport lifetime ($\tau^2$),
due to an interplay of the SOI and broken inversion symmetry. 
Therefore, it can produce larger spin current compared with previous methods. 
This effect is supported by nontrivial spin texture in energy bands that appears in inversion broken systems with the SOI,
 e.g., the surface Weyl state of three-dimensional (3D) topological 
insulators (TI) and two-dimensional (2D) semiconductors with 
the Rashba and/or Dresselhaus SOI.
This new mechanism also offers the rectification of the spin current, i.e., 
the generation of the DC spin current from AC electric fields.
These proposed mechanisms are based on the nonlinear current responses in noncentrosymmetric systems which is captured in the semiclassical treatment using Boltzmann equation as follows.
  
Noncentrosymmetric systems support nonlinear charge current proportional to $E^2$. 
The canonical example is a p-n junction, 
where the difference of $I-V$ characteristics between
the right and left directions leads to the charge current 
proportional to $E^2$. However, for the periodic systems with 
conserved crystal momentum $\bm{k}$, the situation is less trivial.
This is because the time-reversal symmetry $\mathcal{T}$ 
imposes the condition on the energy dispersion, i.e.,
 $\varepsilon_\alpha(\bm{k})= \varepsilon_{\bar{\alpha}}(-\bm{k})$
with $\bar{\alpha}$ being the opposite spin to $\alpha$.
Therefore, even with the broken inversion symmetry $\mathcal{I}$,
there remains a certain symmetry between $\bm{k}$ and $-\bm{k}$
as long as one is concerned about the charge degrees of freedom.
Thus, in Boltzmann transport phenomena where the
charge current is determined by the energy dispersion only,
it is necessary to further break the time reversal symmetry in addition to $\mathcal{I}$, e.g.,  by 
the external magnetic field $\bm{B}$ or the 
spontaneous magnetization $\bm{M}$,
in order to to realize the nonreciprocal charge responses 
\cite{Rikken2001,Krstic2002,Rikken2005,Pop2014,Morimoto2016a,Yasuda2016}. 
Exceptions necessarily require that the
information of the wave functions enters into the transport properties
through e.g. the Berry phase \cite{Sodemann2015,Morimoto2016b}. However, it should be noted that these Berry phase
contributions are not the leading order effect in semiclassics. 
Namely, the dominant one, which is proportional to 
$(\tau E)^2$  in the clean limit, is the contribution captured by the Boltzmann equation.

On the other hand, the situation is dramatically different for the spin current. 
In this case, one needs to distinguish the spin components of the energy bands. 
The spin split bands in noncentrosymmetric systems with the SOI
could produce the spin current proportional to $(\tau E)^2$ even without
breaking the $\mathcal{T}$ symmetry. The difference of the required symmetry for the charge current and the spin current is discussed in detail in the section~\ref{symmetry}.
Since this effect arises from the Boltzmann transport, the generated nonlinear spin current becomes very large (with $\propto \tau^2$) compared with previous methods mentioned above.

We note that the nonlinear spin current in transition metal dichalcogenides (TMDs) 
was also studied theoretically \cite{Yu2014}.
While ref.~\cite{Yu2014} is focused on the band structure with the Ising-type spin splitting along the fixed ($z$-) direction,
our theory is applicable to cases with general SOIs that lacks the $S_z$ conservation. Especially, Rashba system, being intensively studied in the context of the spintronics, is a typical example that breaks $S_z$ conservation. Considering the ubiquitousness of the Rashba system which emerges universally at interfaces and even in the bulk\cite{Ishizaka2011,Sakano2013}, the applicability to such system is a great advantage of the present study for future spintronics studies. 
Furthermore, the nonlinear spin current in the present study is $2$ or $3$ orders of magnitude larger compared with ref.~\cite{Yu2014} since the latter is proportional to a small higher order coefficient, namely, the trigonal warping. The detailed comparison to ref.~\cite{Yu2014} is discussed in the section~\ref{discussion}.
The present nonlinear spin current also ensures controllability of the spin polarization of the flowing spin current through the direction of the electric field and/or the Rashba-Dresselhaus ratio.

\section{Theoretical methods}
\subsection{Boltzmann equation}
First we derive the general formula for nonlinear spin current in the semiclassical regime by using Boltzmann equation.
We consider a system with the electric field $E$ applied in the $x$ direction. 
The Boltzmann equation for the distribution function $f$ is given by
\begin{equation}
-eE\p{f}{k_x} = -\frac{f-f_0}{\tau},
\end{equation}
in the relaxation time approximation ($\tau$ being the relaxation time of electron),
where $f_0$ is the original distribution function in the absence of $E$.
(We have set $\hbar=1$ and adopt the 
convention $e>0$ throughout this paper.)
In order to study the (nonlinear) current response in each order in $E$,
we expand the distribution function as $f=f_0+f_1+f_2+...$, 
where $f_n\propto E^n$ .
The iterative substitution in the
Boltzmann equation yields 
$
f_n=\left( e \tau E \pa{k_x}  \right)^n f_0
$ \cite{Sodemann2015,Yasuda2016,Morimoto2016b,Yu2014}.
In particular, the 
distribution function of the first order in $E$ is given by
\begin{equation}\label{f1}
f_{1}=e\tau E\frac{\partial {f_{0}}}{\partial {k_{x}}}=e\tau E\frac{\partial 
{\varepsilon }}{\partial {k_{x}}}\frac{\partial {f_{0}}}{\partial {\varepsilon }}, 
\end{equation}
and that of the second order in $E$ is \cite{Sodemann2015,Yu2014}
\begin{equation}\label{f2}
f_{2}=e\tau E\frac{\partial {f_{1}}}{\partial {k_{x}}}
=(e\tau E)^2\frac{\partial^2 {f_{0}}}{\partial {k_{x}}^2}.
\end{equation}
The second order term $f_2$ typically shows modulation of electron occupation having the quadrupole 
structure as illustrated in Fig.~\ref{FigTI}(a).

\subsection{definition of spin current}
\begin{figure}[!t]
\centerline{\includegraphics[width=0.5\textwidth]{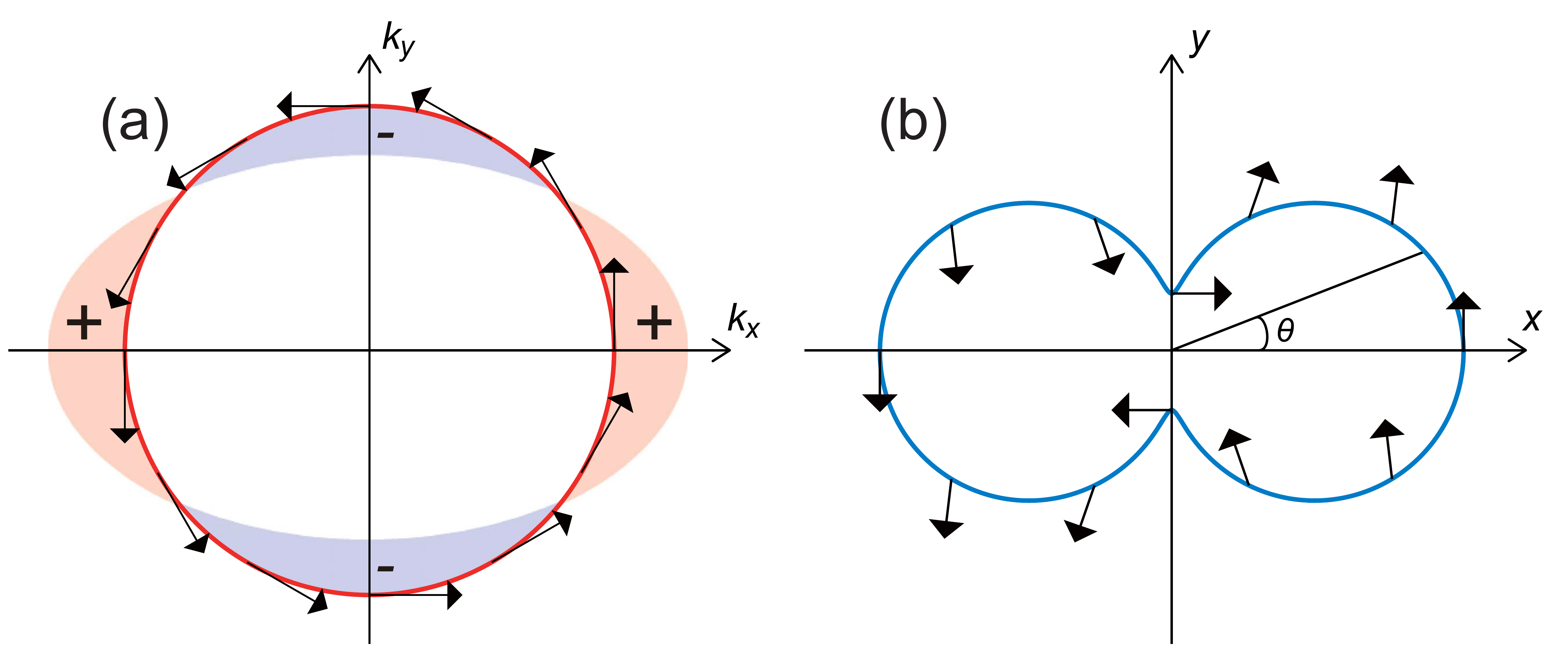}}
\caption{
Second order distribution function and resultant spin current in 3D TI. 
(a) Spin texture along the Fermi surface of the surface state of the 3D TI. 
The shading shows the schematic image of the distribution function of the second 
order in $E$, which is along the $x$ direction. 
We note that this is a schematic picture to clarify the Fermi surface distortion, and the realistic situation for TI with $\tau \sim 1$ ps, $E \sim 1$ kV/m, $v_F\sim 10^5$ m/s, $\mu\sim 10$ meV leads to the distortion of the order of $10^{-4}$ of the Fermi wavenumber.
(b) Spin directions of the spin current 
on the surface of the TI with a parabolic dispersion. The blue curve indicates 
the magnitude, while the arrows show the spin polarization direction of the 
spin current at each direction of flow.
The angle $\theta$ in Fig.1(b) corresponds to the one in eq.(\ref{theta}) in the main text.
}
\label{FigTI}
\end{figure}

The conventional definition of the spin current operator is given by 
the anticommutator of the velocity 
($\propto\frac{\partial \mathcal{\hat{H}}}{\partial k_{\mu }}$) 
and the spin ($\propto\sigma _{\nu})$, $
\hat{\jmath}_{\mu s_{\nu }}\equiv \frac{1}{4}\left\{ \frac{\partial 
\mathcal{\hat{H}}}{\partial k_{\mu }},\sigma _{\nu}\right\}
$~\cite{Murakami2004,Murakami2004a,Sinova2004}.
Hence, the spin current of the $n$th order in $E$ is given by
\begin{equation}
j_{\mu ,s_{\nu }}^{(n)}=\sum_{I }\int 
\frac{\mathrm{d}^{2}\bm{k}}{(2\pi )^{2}} 
\left\langle I,\bm{k}\left\vert \hat{\jmath}_{\mu s_{\nu}}\right\vert I,
\bm{k} \right\rangle f_{n}^{I },
\end{equation}
where $\mu $ is the direction of flow, $\nu $ is the direction of the spin polarization, 
$I$ is the band index, $f_{n}^{I}$ is the $n$th order distribution function for $I$th band.
In the following, we focus on the second order nonlinear spin current $j_{\mu ,s_{\nu }}^{(2)}$ that appears in noncentrosymmetric systems.
Intuitively, an interplay of quadrupole modulation of $f_2$ and nontrivial spin texture due to the SOI [as illustrated in Fig.~\ref{FigTI}(a)] leads to the nonlinear spin current $j_{\mu ,s_{\nu }}^{(2)}$ such as shown in Fig.~\ref{FigTI}(b) as we will see in detail in the section~\ref{TI}.

\section{symmetry argument}\label{symmetry}
The nonlinear charge and spin current ($j_{\mu,s_0}$ and $j_{\mu, s_\nu}$, respectively) are constrained by the time reversal symmetry $\mathcal{T}$. To see this, we suppose that the Hamiltonian satisfies $\mathcal{H}(\vec{k}, \vec{\sigma}) =\mathcal{H}(-\vec{k}, -\vec{\sigma})$,
and hence, every eigenstate has its time-reversal symmetry partner that carries the opposite momentum and opposite spin. First, the charge current $v_\mu =\frac{\partial \mathcal{\hat{H}}}{\partial k_{\mu }}$ is odd under $\mathcal{T}$ ($\mathcal{T}: v_\mu \rightarrow -v_\mu$) while the spin current is even ($\mathcal{T}: \hat{\jmath}_{\mu s_{\nu }} \rightarrow \hat{\jmath}_{\mu s_{\nu }}$).
Next the distribution functions $f_{n}$ is even for even $n$ and odd for odd $n$, because $f_n = \left(e\tau E\right)^n \frac{\partial^n f_0}{\partial k_\mu^n}  \sim (v_\mu)^n$. Therefore, it follows that all odd orders of the  spin current are zero and that all even orders of the  charge-current are zero in the presence of the time-reversal symmetry:
\beq
j^{\text{odd}}_{\mu, s_\nu}  &=& \int \frac{\mathrm{d}^{2}\bm{k}}{(2\pi )^{2}} \ \hat{\jmath}_{\mu s_{\nu }} f_{\text{odd}}= 0, \quad\text{ (with $\mathcal{T}$)},\\
j^{\text{even}}_{\mu, s_0}  &=& \int \frac{\mathrm{d}^{2}\bm{k}}{(2\pi )^{2}} \ \hat{\jmath}_{\mu s_{0 }} f_{\text{even}} = 0, \quad\text{ (with $\mathcal{T}$)}.
\eeq
In particular, we find that the second order charge current vanishes while the second order spin current can be nonvanishing.
Finally, a similar argument applies when a system has the inversion symmetry $\mathcal{I}$ with $\mathcal{H}(\vec{k}, \vec{\sigma}) =\mathcal{H}(-\vec{k}, \vec{\sigma})$.  Since the spin direction is not flipped by the inversion operator (and hence, $\mathcal{I}: \hat{\jmath}_{\mu s_{\nu }} \rightarrow -\hat{\jmath}_{\mu s_{\nu }}$), all charge and spin nonlinear current in the even order are zero:
\beq
j^{\text{even}}_{\mu, s_\nu}  &=& \int \frac{\mathrm{d}^{2}\bm{k}}{(2\pi )^{2}} \ \hat{\jmath}_{\mu s_{\nu }}f_{\text{even}}= 0, \quad\text{ (with $\mathcal{I}$)},\\
j^{\text{even}}_{\mu, s_0}  &=& \int \frac{\mathrm{d}^{2}\bm{k}}{(2\pi )^{2}} \ \hat{\jmath}_{\mu s_{0 }} f_{\text{even}} = 0, \quad\text{ (with $\mathcal{I}$)}.
\eeq
These symmetry analyses indicate that the nonlinear spin current $\propto E^2$ in the Boltzmann transport requires broken inversion, but it does not require broken time-reversal symmetric systems.
In the following sections, we study a few examples of noncentrosymmetric systems with the SOI that support the nonlinear spin current.

\section{Surface state of the 3D TI}\label{TI}
We start with the surface of a 3D TI. It is described by the Hamiltonian
$
\hat{\mathcal{H}} _{\text{TI}}= v (k_x\sigma_y - k_y \sigma_x),
$
where $v$ is the velocity of the Weyl cone. The energy dispersion is 
$\varepsilon^I=Ivk$ with $I=\pm$, and the spin polarization for each 
branch in the $k$ space is $\left\langle \pm,\bm{k} 
\left\vert \vec{\sigma}\right\vert \pm,\bm{k}
\right\rangle  = \pm( -\sin\phi,\cos\phi,0)$, where $k_x=k\cos\phi , k_y=k\sin\phi$. 
We show the Fermi surface (FS) and the spin direction for the upper branch together 
with the second order distribution function in Fig.~\ref{FigTI}(a). By using spin current 
operators,
$
\hat{\jmath}_{xs_{x}}=\hat{\jmath}_{ys_{y}}=0
$
and
$
\hat{\jmath}_{xs_{y}}=-\hat{\jmath}_{ys_{x}}=\frac{1}{2}v,
$
we can show that
$
j_{\mu ,s_{\nu }}^{(n)} \propto \int 
\frac{\mathrm{d}^{2}\bm{k}}{(2\pi )^{2}}\frac{\partial
^{n}f_{0}^{\pm }}{\partial k_{x}^{n}}=0.
$
Namely, all the spin currents are zero.

However, nonzero spin currents are generated in the presence of the parabolic term
$k^2/(2m)$ in the Hamiltonian $\hat{\mathcal{H}} _{\text{TI}}$;
\begin{equation}
 \hat{\mathcal{H}} _{\text{TI}} = \frac{k^2}{2m} + v (k_x\sigma_y - k_y \sigma_x).
\end{equation}
The emergence of the parabolic dispersion is expected in general when the system has a 
band asymmetry between the electron and hole bands.

The energy dispersion is given by 
\begin{equation}
\varepsilon^\pm (\kk ) = \frac{k^2}{2m} \pm vk
\end{equation}
and the Fermi surface is formed by one of these two branches depending on the sign of the chemical potential $\mu$. The Fermi momentum is determined as $k_F^\pm =\mp mv\pm\sqrt{2m\mu+ m^2v^2}$ with $\pm$ corresponding to the sign of $\mu$.
The velocity operators in this case are given as
\begin{equation}
\frac{\partial \mathcal{\hat{H}}}{\partial k_{x }} = \frac{k_x}{m} + v\sigma_y, \qquad  \frac{\partial \mathcal{\hat{H}}}{\partial k_{y }} = \frac{k_y}{m} -v\sigma_x.
\end{equation}
The spin current operators are given by 
\begin{eqnarray}
\hat{\jmath}_{xs_{x}} &=&\frac{k_x}{2m}\sigma_x, \qquad
\hat{\jmath}_{xs_{y}} =\frac{k_x}{2m}\sigma_y+\frac{1}{2}v, \nonumber \\
\hat{\jmath}_{ys_{x}} &=&\frac{k_y}{2m}\sigma_x-\frac{1}{2}v, \qquad
\hat{\jmath}_{ys_{y}} =\frac{k_y}{2m}\sigma_y,
\end{eqnarray}
which are summarized as $\hat{\jmath}_{\mu s _{\nu}} = \frac{k_\mu}{2m}\sigma_{\nu} $ up to irrelevant constant terms. And their expectation values for each branch of Weyl cone are
\begin{eqnarray}
\left\langle \pm ,\boldsymbol{k} \left\vert \hat{\jmath}_{xs_{x}}\right\vert \pm ,\boldsymbol{k} \right\rangle  &=&\mp \frac{1}{2}\frac{k}{m}\sin \phi \cos \phi , \nonumber \\
\left\langle \pm ,\boldsymbol{k} \left\vert \hat{\jmath}_{xs_{y}}\right\vert \pm ,\boldsymbol{k} \right\rangle  &=&\pm \frac{k}{2m}\cos ^{2}\phi +\frac{1}{2}v  , \nonumber \\
\left\langle \pm ,\boldsymbol{k} \left\vert \hat{\jmath}_{ys_{x}}\right\vert \pm ,\boldsymbol{k} \right\rangle  &=&\mp \frac{k}{2m}\sin ^{2}\phi -\frac{1}{2}v  , \nonumber \\
\left\langle \pm ,\boldsymbol{k} \left\vert \hat{\jmath}_{ys_{y}}\right\vert \pm ,\boldsymbol{k} \right\rangle  &=&\pm \frac{1}{2}\frac{k}{m}\sin \phi \cos \phi .
\end{eqnarray}

As expected from the symmetry argument, all the linear spin currents vanish after the $\phi $ integration;
$ j_{x,s_{x}}^{(1)}=j_{x,s_{y}}^{(1)}=j_{y,s_{x}}^{(1)}=j_{y,s_{y}}^{(1)}=0. $ This result can be shown explicitly as follows. All the expectation values of the spin currents are the zeroth or the second order in $\cos\phi$ or $\sin\phi$ while $f_1\propto\cos\phi$. The product of these two terms are first or third order in $\cos\phi$ or $\sin\phi$ which vanishes by the $\phi $ integration.

Second order spin currents are calculated by the integration by part at zero temperature as
\begin{eqnarray}
j_{x,s_{y}}^{(2)} &=&\int \frac{\td^{2}\boldsymbol{k}}{{(2\pi )}^{2}}\ 
\left\langle \pm ,\boldsymbol{k} \left\vert \hat{\jmath}_{xs_{y}}\right\vert
\pm ,\boldsymbol{k} \right\rangle f_{2}^{\pm }  \nonumber \\
&=&\int \frac{\td^{2}\boldsymbol{k}}{{(2\pi )}^{2}}\ \left[ \frac{1}{2}\left( \pm \frac{k}{m}\cos ^{2}\phi +v \right) \right]  \nonumber\\
& & \qquad \qquad  \times \left[ e^{2}\tau ^{2}E^{2}\frac{\partial }{\partial k_{x}} \frac{\partial \varepsilon ^\pm}{\partial k_x} \df \right]  \nonumber \\
&=&\mp \frac{e^{2}\tau ^{2}E^{2}}{8\pi ^{2}m}\int \td^{2}\boldsymbol{k}\ 
\left[ \cos ^{3}\phi +2\sin^2\phi \cos \phi )\right] \nonumber\\
& & \qquad \qquad  \times \left[ \left( \frac{k}{m}\pm v \right) \cos \phi \df \right]  \nonumber \\
&=&\mp \frac{5e^{2}\tau ^{2}E^{2}}{32\pi m}\int k \td k \ 
 \left( \frac{k}{m}\pm v \right) \df  \nonumber \\
&=& \frac{5e^{2}\tau ^{2}E^{2}}{32\pi m}\int k \td k \ \delta (k-k_F^\pm )\nonumber \\
&=& \pm \frac{5e^{2}\tau ^{2}E^{2}}{32\pi m} \left[ - mv + \sqrt{2m\mu+ m^2v^2} \right]. \label{js2TI}
\end{eqnarray}
By similar calculation shown in Appendix, we have $j_{y,s_{x}}^{(2)}=\frac{1}{5}j_{x,s_{y}}^{(2)}$ and $j_{x,s_{x}}^{(2)}=j_{y,s_{y}}^{(2)}=0$. Note that the signs of the spin currents depend on the sign of chemical potential $\mu$.

Nonzero spin current generation is naturally understood in terms of the spin direction at the 
FS and the second-order distribution function possessing a quadrupole structure: See Fig.
\ref{FigTI}(a). Namely, the distribution function is positive toward $\pm x$ direction and 
hence both the $+y$ spin flowing in the $+x$ direction and the $-y$ spin flowing 
in the $-x$ direction are accelerated by the application of $E$ parallel to the $x$ direction. 
In total, $j_{x,s_{y}}^{(2)}$ becomes positive. Similarly, since $f_2$ is negative toward 
$\pm y$ direction, both the $-x$ spin flowing in the $+y$ direction and the $+x$ spin 
flowing in the $-y$ direction are negatively accelerated, thus resulting in the positive 
$j_{y,s_x}^{(2)}$.

In order to clarify the real space texture of the generated spin current, we define the 
spin current toward the $\theta$ direction as
\begin{eqnarray}
\vec{j}^{(2)}_{\theta s}  \equiv \left(
\begin{array}{l}
j^{(2)}_{\theta s_x} \\
j^{(2)}_{\theta s_y} \\
\end{array}
\right)  \equiv \left(
\begin{array}{l}
 j^{(2)}_{xs_x}\cos\theta +j^{(2)}_{ys_x}\sin\theta \\
j^{(2)}_{xs_y}\cos\theta +j^{(2)}_{ys_y} \sin\theta \\
\end{array}
\right) . \label{theta}
\end{eqnarray}
We show the polar plot of $\vec{j}^{(2)}_{\theta s} $ in Fig.~\ref{FigTI}(b), where 
the blue line shows the amplitude of the spin current $|\vec{j}^{(2)}_{\theta s}| $ 
while the black arrows show the direction of the spin polarization $\vec{j}^{(2)}_{\theta s} / 
|\vec{j}^{(2)}_{\theta s}| $. Using the fact $j^{(2)}_{xs_y}=5j^{(2)}_{ys_x}$ 
and $j^{(2)}_{xs_x}=j^{(2)}_{ys_y}=0$, the magnitude of the spin current  is given 
by 
$|\vec{j}^{(2)}_{\theta s}| \equiv \sqrt{(j^{(2)}_{\theta s_x})^2
+(j^{(2)}_{\theta s_y})^2} = |j^{(2)}_{ys_x}|\sqrt{25\cos^2\theta+\sin^2\theta}$, 
which well describes the blue curve in Fig.~\ref{FigTI}(b).

\begin{figure}[!t]
\centerline{\includegraphics[width=0.5\textwidth]{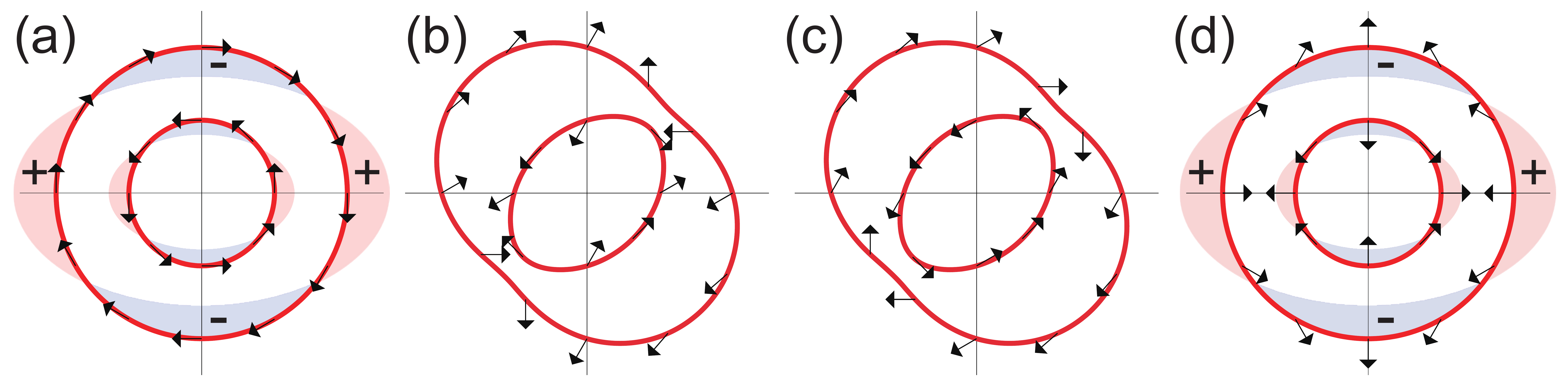}}
\caption{
Spin textures of the Fermi surfaces of the Rashba-Dresselhaus system. 
(a) For $\tan^{-1}(\beta/\alpha)=0$ (Rashba system). (b)For  $\pi/6$. (c) For $\pi/3$. (d) For $\pi/2$ (Dresselhaus system). The shading shows the schematic image of the 
distribution function of the second order in $E$.
Here, we have set $\mu =0.2$, $m=1$ and $\alpha^2+\beta^2 =1$.}
\label{FigFS}
\end{figure}

\section{Rashba-Dresselhaus system}
  Rashba and Dresselhaus type SOIs are present in wide classes of materials without inversion symmetry.
 The Hamiltonian including the Rashba type and the linear Dresselhaus type SOIs is
\begin{equation}
\hat{\mathcal{H}}_{\text{RD}} = \frac{k^2}{2m} + 
\alpha (k_x\sigma_y - k_y \sigma_x)+\beta \left( k_{x}\sigma _{x}-k_{y}
\sigma _{y}\right) , 
\end{equation}
where $m$ is the electron effective mass, $\alpha$ is the Rashba SOI strength and 
$\beta$ is the Dresselhaus SOI strength.
There are two bands indexed by $I=\pm$,
\begin{equation}
\varepsilon ^\pm\left( \bm{k}\right) =\frac{k^{2}}{2m}\pm k\sqrt{\alpha^{2}
+\beta ^{2}-2\alpha \beta \sin 2\phi }.
\end{equation}
The spin polarization in the $k$ space is  $\left\langle \pm,\bm{k} \left\vert 
\vec{\sigma}\right\vert \pm,\bm{k} \right\rangle  = \pm( \cos\varphi,-\sin\varphi,0)$
, where $\varphi\equiv\mathrm{arg} \left[(\beta k_x-\alpha k_y) 
+i(\beta k_y-\alpha k_x) \right]$.
We show FSs and the spin textures 
for various values of $\tan^{-1}(\beta /\alpha)$ 
in Fig.~\ref{FigFS} while keeping $\alpha^2 + \beta^2=1$. FSs are anisotropic for the general Rashba-Dresselhaus system. 
In this case there are two FSs in contrast to the case of the surface state of TI.

The anisotropic Fermi momentum for the upper band is
$k_{F+}^+ = - mA + \sqrt{m^2A^2+2m\mu},$
while those for lower bands are
$k_{F-}^\pm = + mA \pm \sqrt{m^2A^2+2m\mu}$,
with $A(\phi )=\sqrt{\alpha^2+\beta^2 -2\alpha\beta \sin2\phi}$.
For $\mu>0$, $k_{F+}^+$ and $k_{F-}^+$ form Fermi surfaces, while $k_{F-}^+$ and $k_{F-}^-$ do for $\mu<0$.
Note that the Fermi surface for $\mu<0$ vanishes for $\phi$ such that $m^2A(\phi)^2+2m\mu < 0$.

Velocity operators are given as
\begin{equation}
\p{\mathcal{\hat{H}}}{k_x} = \frac{k_x}{m} + \alpha\sigma_y+\beta\sigma_x , \quad \p{\mathcal{\hat{H}}}{k_y} = \frac{k_y}{m} - \alpha\sigma_x-\beta\sigma_y.
\end{equation}
From these, we have spin current operators as
\begin{eqnarray}
\hat{j}_{x s_x} &=& \frac{1}{2}  \left(\frac{k_x}{m}\sigma_x +\beta \right), \quad 
\hat{j}_{x s_y} = \frac{1}{2}  \left( \frac{k_x}{m}\sigma_y +\alpha \right),\nonumber\\
\hat{j}_{y s_x} &=& \frac{1}{2}  \left( \frac{k_y}{m}\sigma_x -\alpha \right), \quad
\hat{j}_{y s_y} = \frac{1}{2}  \left( \frac{k_y}{m}\sigma_y -\beta \right),
\end{eqnarray}
which are again summarized as $\hat{\jmath}_{\mu s _{\nu}} = \frac{k_\mu}{2m}\sigma_{\nu} +const.$. And their expectation values for each band are
\begin{eqnarray}
\left\langle \pm ,\boldsymbol{k} \left\vert \hat{\jmath}_{xs_{x}}\right\vert \pm ,\boldsymbol{k} \right\rangle &=&   \frac{1}{2}  \left( \pm\frac{k}{m} \cos \phi\cos\varphi +\beta \right),  \nonumber\\
\left\langle \pm ,\boldsymbol{k} \left\vert \hat{\jmath}_{xs_{y}}\right\vert \pm ,\boldsymbol{k} \right\rangle &=&  \frac{1}{2}  \left( \mp\frac{k}{m} \cos \phi \sin\varphi +\alpha \right),  \nonumber\\
\left\langle \pm ,\boldsymbol{k} \left\vert \hat{\jmath}_{ys_{x}}\right\vert \pm ,\boldsymbol{k} \right\rangle&=&   \frac{1}{2}  \left( \pm\frac{k}{m} \sin\phi \cos\varphi -\alpha \right),  \nonumber\\
\left\langle \pm ,\boldsymbol{k} \left\vert \hat{\jmath}_{ys_{y}}\right\vert \pm ,\boldsymbol{k} \right\rangle &=&   \frac{1}{2}  \left( \mp\frac{k}{m} \sin\phi \sin\varphi -\beta\right).
\end{eqnarray}

\begin{figure}[!t]
\centerline{\includegraphics[width=0.5\textwidth]{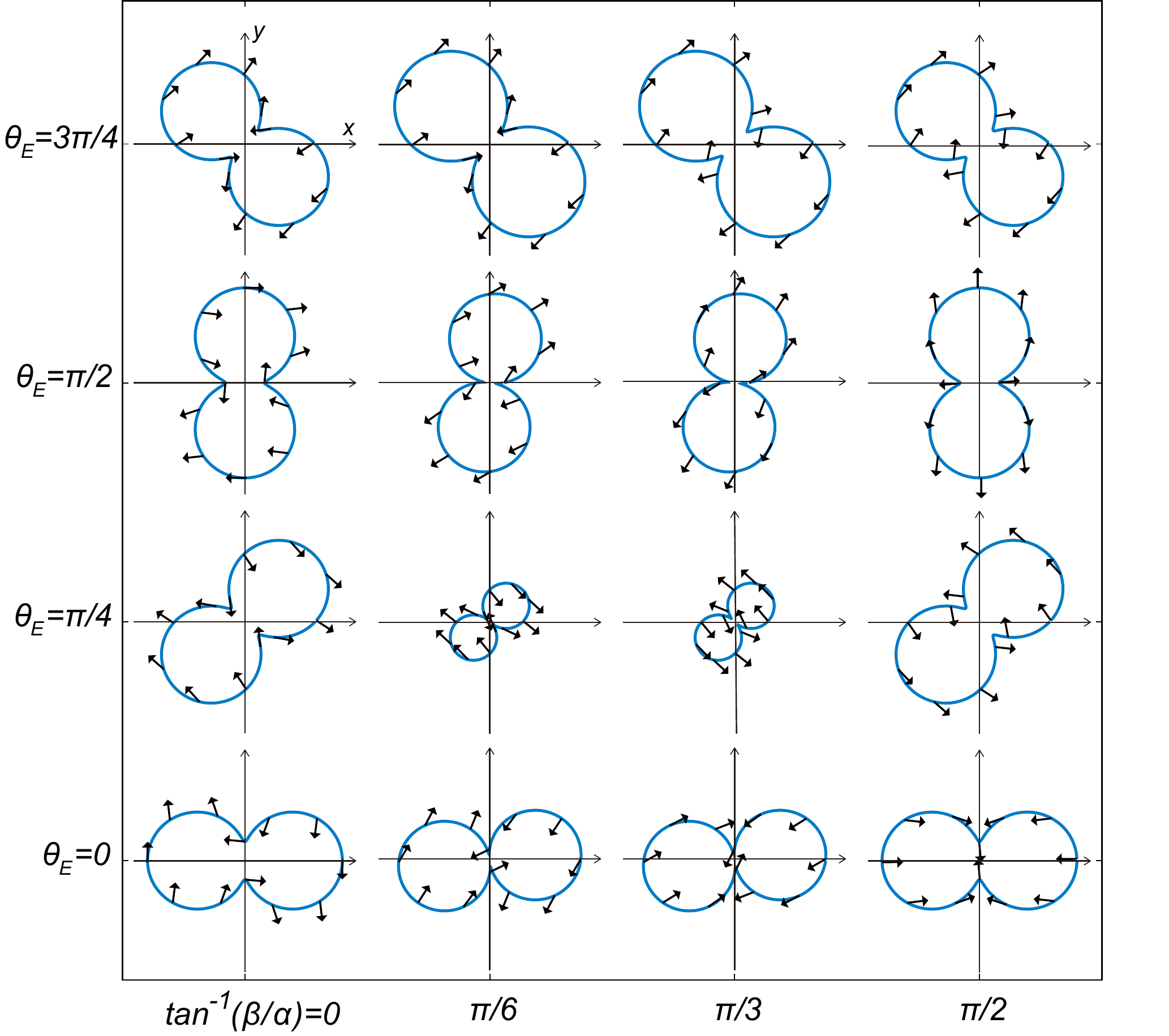}}
\caption{
Spin polarization directions of the spin current in the Rashba-Dresselhaus system with 
general interaction strength and the electric field direction.
The blue curves show the magnitude of the spin current, while the vectors show 
the direction of the spin polarization. Horizontal axis is Rashba-Dresselhaus ratio $\tan^{-1}(\beta /\alpha )$ and vertical is the electric field direction $\theta_{E} $ where $\bm{E} = (E \cos \theta_E, E \sin \theta_E)$. 
Here, we have set $\mu =0.2$, $m=1$ and $\alpha^2+\beta^2 =1$.}
\label{FigJR}
\end{figure}

Using these results, we numerically calculated the second-order spin current for some values of 
$\tan^{-1}(\beta /\alpha)$ and the direction of the applied electric field $\theta_{E}$, 
where $\bm{E}=E(\cos\theta_E,\sin\theta_E)$. The polar plot of the spin current 
is summarized in Fig.~\ref{FigJR}. Note that the distribution function under the 
application of the electric field in general direction is obtained by a simple substitution 
$E\frac{\partial}{\partial k_x} \rightarrow \bm{E}\cdot 
\frac{\partial}{\partial \bm{k}}$ in eqs. (\ref{f1}) and (\ref{f2}).
We have numerically confirmed that all the first-order spin currents are zero, which is consistent with the symmetry requirement. 
We have also confirmed the chemical potential dependence is negligible when $\mu>0$. 
Detailed arguments for the Rashba system ($\tan^{-1}(\beta /\alpha)=0$, the 
leftmost column in Fig.~\ref{FigJR}) and the Dresselhaus system 
($\tan^{-1}(\beta /\alpha)=\pi/2$ , the rightmost column in Fig.~\ref{FigJR}) 
are given below.

\subsection{Rashba system}
 We first investigate the pure Rashba system, 
for which $\alpha\neq 0$, $\beta =0$  and $\theta_E=0$.
The eigenstates and the spin polarization are the same as those in the 
surface state of 3D TI; $ \left\langle \pm,\bm{k} \left\vert 
\vec{\sigma}\right\vert \pm,\bm{k} \right\rangle  = \pm( \cos\varphi,-\sin\varphi,0) =\pm( -\sin\phi,\cos\phi,0). $
We show the spin textures of the FSs in the pure 
Rashba system in Fig.~\ref{FigFS}(a).
The spin texture forms vortex structures, 
whose directions are opposite between the 
inner and outer FSs.
All the first-order spin currents are analytically shown to vanish by the $\phi $ integration; 
$j_{x,s_{x}}^{(1)R}=j_{x,s_{y}}^{(1)R}=j_{y,s_{x}}^{(1)R}
=j_{y,s_{y}}^{(1)R}=0$ where the superscript $R$ indicates Rashba system.  This is consistent with the symmetry argument. Furthermore, second-order spin currents are calculated as
\begin{eqnarray}
j_{x,s_{y}}^{(2)R} &=&\sum_{\pm}\int \frac{\td^{2}\boldsymbol{k}}{{(2\pi )}^{2}}\ 
\left\langle \pm ,\boldsymbol{k} \left\vert \hat{\jmath}_{xs_{y}}\right\vert
\pm ,\boldsymbol{k} \right\rangle f_{2}^{\pm }  \notag \\
&=&\sum_{\pm}\int \frac{\td^{2}\boldsymbol{k}}{{(2\pi )}^{2}}\ 
\left[ \frac{1}{2}\left( \pm \frac{k}{m}\cos ^{2}\phi +\alpha \right) \right] \nonumber \\
& & \qquad\qquad \times \left[ e^{2}\tau ^{2}E^{2}\frac{\partial }{\partial k_{x}}\left( \frac{k}{m}\pm \alpha \right) 
\cos \phi \df
\right]  \notag \\
&=&\sum_{\pm}\mp \frac{e^{2}\tau ^{2}E^{2}}{8\pi ^{2}m}\int \td^{2}\boldsymbol{k}\ 
\left[ \cos ^{3}\phi +2\sin^2\phi \cos \phi )\right] \nonumber\\
& & \qquad\qquad \times\left[ \left( \frac{k}{m}\pm \alpha \right) \cos
\phi \df\right]  \notag \\
&=&\sum_{\pm}\mp \frac{5e^{2}\tau ^{2}E^{2}}{32\pi m}\int k \td k \ 
 \left( \frac{k}{m}\pm \alpha \right) \df \notag \\
&=& \frac{5e^{2}\tau ^{2}E^{2}}{32\pi m}\times\left\{ 
\begin{array}{ll}
k_{F+}^+ -k_{F-}^+  & (\mu >0) \\ 
-k_{F-}^+ +k_{F-}^- & (\mu <0)
\end{array}
\right. \notag\\
&=&- \frac{5e^{2}\tau ^{2}E^{2}}{16\pi m}\times\left\{ 
\begin{array}{ll}
m\alpha & (\mu >0) \\ 
\sqrt{2m\mu+m^2\alpha^2}& (\mu <0).
\end{array}
\right. \label{js2rashba}
\end{eqnarray}

Similarly,  we have $j_{y,s_{x}}^{(2)R}=\frac{1}{5}j_{x,s_{y}}^{(2)R}$ and $j_{x,s_{x}}^{(2)R}=j_{y,s_{y}}^{(2)R}=0$
This relation is the same as that in the case of the TI.
We note that the sign of the spin currents is opposite compared to that in the TI.

The polar plot of the second-order spin current in the Rashba system is shown in 
Fig.~\ref{FigJR} in the panel corresponding to $\tan^{-1}(\beta /\alpha )=0$ and 
$\theta_{E}=0$. The shape of the pattern is completely the same as that in 3D TI 
(Fig.~\ref{FigTI}(b)) but the spin polarization is opposite.

\subsection{Dresselhaus system}
 We next investigate the Dresselhaus system, 
for which $\beta\neq 0$, $\alpha =0$ and $\theta_E=0$. We show the spin direction of the FSs
in the pure Dresselhaus system in Fig.~\ref{FigFS}(d). The spin texture
forms hedgehog structures, whose directions are opposite between the inner 
and outer FSs;  $\left\langle \pm,\bm{k} \left\vert \vec{\sigma}\right\vert 
\pm,\bm{k} \right\rangle  = \pm( \cos\varphi,-\sin\varphi,0) = \pm( \cos\phi,-\sin\phi,0)$.
The eigenenergy and the distribution functions between the Rashba Hamiltonian and 
the Dresselhaus Hamiltonian are the same. Only the difference is the expectation value of the spin current operators. 

We find the relation between expectation values of spin current operators of the Rashba and the Dresselhaus systems as
\begin{eqnarray}
\left. \left\langle \pm ,\boldsymbol{k} \left\vert \hat{\jmath}_{xs_{x}}^{D}\right\vert \pm ,\boldsymbol{k} 
\right\rangle  _D \right| _{\beta\rightarrow\alpha}&=&\left\langle \pm ,\boldsymbol{k} \left\vert \hat{\jmath}_{xs_{y}}^{R}\right\vert \pm ,\boldsymbol{k} \right\rangle _R , \\
\left. \left\langle \pm ,\boldsymbol{k} \left\vert \hat{\jmath}_{xs_{y}}^{D}\right\vert \pm ,\boldsymbol{k}
\right\rangle _D \right| _{\beta\rightarrow\alpha}&=&\left\langle \pm  ,\boldsymbol{k}\left\vert \hat{\jmath}_{xs_{x}}^{R}\right\vert \pm  ,\boldsymbol{k}\right\rangle_R , \\
\left. \left\langle \pm  ,\boldsymbol{k}\left\vert \hat{\jmath}_{ys_{x}}^{D}\right\vert \pm ,\boldsymbol{k}
\right\rangle _D \right| _{\beta\rightarrow\alpha}&=&\left\langle \pm  ,\boldsymbol{k}\left\vert \hat{\jmath}_{ys_{y}}^{R}\right\vert \pm  ,\boldsymbol{k}\right\rangle_R , \\
\left. \left\langle \pm  ,\boldsymbol{k}\left\vert \hat{\jmath}_{ys_{y}}^{D}\right\vert \pm ,\boldsymbol{k}
\right\rangle  _D \right| _{\beta\rightarrow\alpha}&=&\left\langle \pm  ,\boldsymbol{k}\left\vert \hat{\jmath}_{ys_{x}}^{R}\right\vert \pm  ,\boldsymbol{k}\right\rangle_R ,
\end{eqnarray}
where super and subscripts $R/D$ denote the Rashba/Dresselhaus systems. These relations and the equivalence of the band dispersion guarantee all the linear spin currents to be zero as expected. Furthermore, the second-order spin currents are given by 
$
j_{x,s_{x}}^{(2)\text{D}}=5 j_{y,s_{y}}^{(2)\text{D}}=j_{x,s_{y}}^{(2)\text{R}}=5j_{y,s_{x}}^{(2)\text{R}},
$
and $j_{x,s_{y}}^{(2)\text{D}}=j_{y,s_{x}}^{(2)\text{D}}=0$ where superscripts $D$ is for Dresselhaus system. 
The polar plot of the second-order spin current in the Dresselhaus system is shown in 
Fig.~\ref{FigJR}. See the panel corresponding to $\tan^{-1}(\beta /\alpha )=\pi/2$ 
and $\theta_{E}=0$ therein. The peanuts-like shape is completely the same as those 
in 3D TI and the Rashba system, but the spin polarization reflects the hedgehog 
structure at FSs.

\subsection{Carrier density and temperature dependences}

\begin{figure}[!t]
\centerline{\includegraphics[width=0.5\textwidth]{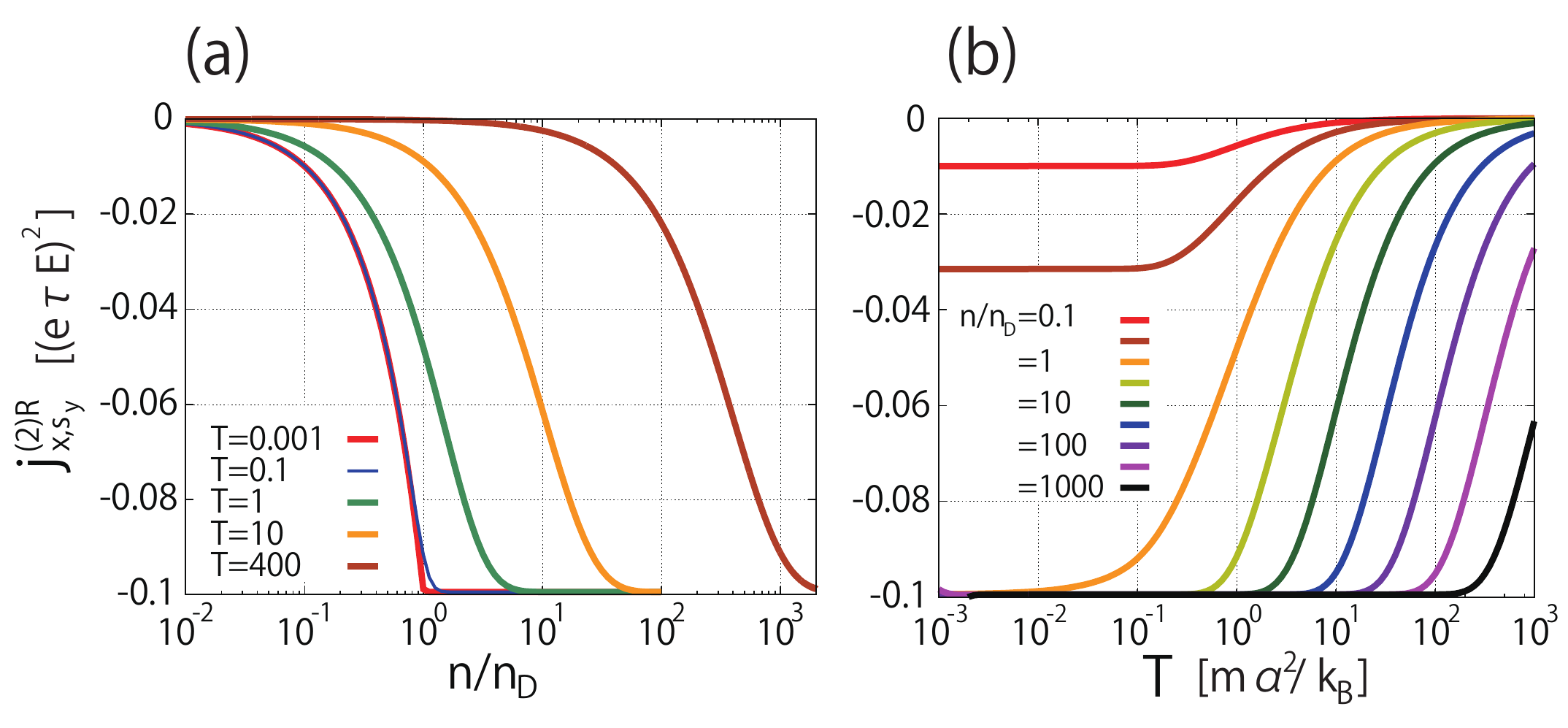}} 
\caption{ The carrier density and the temperature dependence of the second-order spin 
current in the Rashba system $\bm{j_{x,s_{y}}^{(2)\text{R}}}$. 
$n_D=m^2\alpha^2 /\pi$ is the carrier density corresponding to the Dirac point 
at the zero temperature. Here, we have set $m=\alpha=1$.}
\label{FigfiniteT}
\end{figure}

Now we consider the dependence of the spin current on the 
carrier density $n$ and temperature $T$. 
We show the carrier density and temperature dependence of $j_{x,s_{y}}^{(2)\text{R}}
 (=j_{x,s_{x}}^{(2)\text{D}}=5j_{y,s_{x}}^{(2)\text{R}}=5j_{y,s_{y}}^{(2)\text{D}}
)$ in Fig.~\ref{FigfiniteT}.
We take the Rashba system for example here, but
the generic features are common for other cases also. 
Equation (\ref{js2rashba}) and Fig.~\ref{FigfiniteT}(a) indicate that the magnitude of the spin current at the zero temperature increases as the increase of carrier density $n$ and 
becomes constant for $n >n_D$ with $n_D=m^2\alpha^2 /\pi$ being the carrier 
density corresponding to the Dirac point. 
According to eq.(\ref{js2rashba}), the 
magnitude of the spin current is proportional to the difference of the Fermi 
momentum defined for each FS, which is constant above the Dirac point. The constant 
spin current above the Dirac point indicates that the effect of finite temperature is tiny as shown in Fig.~\ref{FigfiniteT}(b).

\section{Discussion} \label{discussion}
We have demonstrated that the spin current
of the second order in $E$ is generated in noncentrosymmeric
systems with nontrivial spin textures in the momentum space. We also note that the amplitude of the spin 
current is $2$ or $3$ orders of magnitude larger than the previous proposal on TMDs\cite{Yu2014}, 
indicating that our mechanism can generate nonlinear spin current more efficiently. In TMDs, the anisotropic Fermi surface due to the trigonal warping plays the crucial role in the spin current generation. Authors of ref.~\cite{Yu2014} claimed that the generated nonlinear spin current normalized by the linear charge current is 
\begin{equation}
\frac{j^{(2)}_s \times 2/\hbar}{j^{(1)}_c /e} = \frac{3\gamma e \tau E }{\hbar} \label{beta}
\end{equation}
where $\gamma$ is the coefficient of the trigonal warping which has the dimension of the length. (See eqs. (3) and (5) in ref.~\cite{Yu2014}. We have replaced the coefficient $\beta$ in ref.~\cite{Yu2014} by $\gamma$ to avoid the confusion. ) The $\gamma$ values are summarized in the Table 1 in ref.~\cite{Yu2014}, which is of the order of $0.1\sim1 \mathrm{\AA}$ for MoS$_2$ and GaSe.
 To show that our proposed method is more efficient mechanism to generate the spin current, we calculated the same ratio for our system and define $\gamma$ parameter by eq.(\ref{beta}). For Rashba system with $\mu >0$ for example, the linear charge current $j_{c}^{(1)}=\frac{e^2\tau E}{2\pi } (2\mu + m\alpha ^2)$ and the second order spin current is given in eq.(\ref{js2rashba}).
The $\gamma$ value is calculated as
$
\gamma = \frac{5 \alpha}{12}\frac{1}{2\mu+m\alpha^2} < \frac{5}{12m\alpha}.
$
The maximum value of this ratio is achieved by setting $\mu=0$. In this case the $\gamma$ value is $630 \mathrm{\AA}$ for GaAs 
by substituting $\alpha\simeq 0.1$ eV\AA \cite{Simmons2015} and $m\simeq 0.3 m_{e}$ \cite{Glover1973}. 
For the bulk Rashba semiconductor BiTeI, $\gamma $ parameter is $8.1 \mathrm{\AA}$ by assuming $\alpha\simeq 3.9\times $ eV\AA \cite{Ishizaka2011} and $m\simeq 0.15 m_{e}$\cite{Sakano2013}. 
Here, $m_e$ is the electron mass in the vacuum. 
For the surface of the TI, the linear charge current is $j_{c}^{(1)}=\frac{e^2\tau E}{4\pi m}\left[-mv+\sqrt{2m\mu+m^2v^2}\right] \sqrt{2m\mu+m^2v^2}$ and second order spin current is given in eq.(\ref{js2TI}).
The $\gamma$ parameter is
$
\gamma = \frac{5 }{12}\frac{1}{\sqrt{2m\mu+m^2v^2}} \sim \frac{5 }{12mv}
$
by assuming $2m\mu \ll m^2v^2$. 
This value is about $17 \mathrm{\AA}$ for 3D TI $\mathrm{Bi_2Se_3}$ by using $v\simeq 6.2\times 10^5$ m/s \cite{Zhang2009} and $m=0.53m_{e}$ \cite{Kim2016}.
These three values are much larger than that discussed in ref \cite{Yu2014}.
Thus, we can conclude that our proposed method is more efficient mechanism to generate the nonlinear spin current.

Generation of the spin current proportional to $E^2$ indicates that the DC spin current is 
induced by the AC electric field $E(t)=Ee^{i\omega t}$. The time-dependent Boltzmann 
equation yields the second order distribution function which is composed of two terms; the 
time-independent term and the one with $2\omega$ frequency \cite{Sodemann2015}. 
The latter one vanishes in time-average, while the former gives us a finite rectified spin 
current, which is calculable by the equivalent procedure as in the present study. This 
rectified spin current can be induced for example by shining the terahertz light.

Under the irradiation of the light on systems with spin-splitted bands, the circular photogalvanic effect also contributes to the spin current associated with the charge current. The interband transition with optical selection rule gives us an unbalanced distribution of the positive and negative momenta on the spin splitted band resulting in the spin polarized photocurrent \cite{Ganichev2001,Ganichev2002a}. Similarly, photocurrent is also generated by spin galvanic effect. The optical spin accumulation by the absorption of the circularly polarized light results in the photocurrent induction in the assymetric spin flip scattering processes \cite{Ganichev2002b,Ganichev2004}. However, these phenomena can be excluded by using the linearly polarized terahertz light which does not selectively excite electrons with lifted spin degeneracy. 

We next estimate the magnitude of the spin current for various systems. We define the 3D spin 
conductivity as $j_{x,s_y}^{(2)}/(E\hbar/(2ec))$, where $c$ is the lattice constant for thickness 
direction and we assume the reasonable value of the magnitude of electric field $E\simeq 10^{2\sim 5}$ V/m. The spin conductivity is the order of $2\times 10^{2\sim 5}\ 
\mathrm{\Omega^{-1}m^{-1}}$ for GaAs by substituting $\tau\simeq 2.5$ ps and $c=5.7\mathrm{\AA}$.
It is also the order of $7\times 10^{0\sim 3} \ \mathrm{\Omega^{-1}m^{-1}}$ for BiTeI with $\tau\simeq 0.072$ ps \cite{Wang2013a} and $c=6.9\mathrm{\AA}$. 
As for 3D TI, it is the order of $1.3\times 10^{2\sim 5} \ 
\mathrm{\Omega^{-1}m^{-1}}$ for Bi$_2$Se$_3$ by substituting $\tau\simeq 3.1$ ps 
\cite{Glinka2013}, $c=29 \mathrm{\AA}$ and $\mu=0.1$ eV.
These values are larger than the typical value of the spin Hall conductivity 
$10^{0\sim 4} \ \mathrm{\Omega^{-1}m^{-1}}$ \cite{Sinova2015}. 
The effect of finite temperature summarized in Fig. \ref{FigfiniteT} has a peculiar feature. 
For the typical sheet carrier density of the order of $\sim10^{13} \mathrm{cm^{-2}}$, 
the carrier density $n/n_D$ is the order of $10^4$ for GaAs and $1$ for BiTeI. The room 
temperature in Fig.~\ref{FigfiniteT}, $300\mathrm{K}/ (m\alpha^2/k_B)$, is about $400$ 
for GaAs and $0.1$ for BiTeI. As seen in Fig.~\ref{FigfiniteT}, we may conclude that the 
spin current never reduces even at room temperature. 

Finally, we discuss the validity of the present work. Our derivation of the second order 
distribution function is based on the expansion with respect to $\tau e E \frac{\partial}{\partial \hbar 
k_x} \sim \tau e E /(\hbar k_0)$, where $k_0 \sim m\alpha/\hbar^2$ is the typical 
momentum of, for instance, the Rashba system. For the convergence of the expansion,  
the electric field $E$ must satisfy $E \ll m \alpha/(e\hbar\tau ) $. This condition has two 
physical interpretations. One is that  the energy due to the electric field $eE/k_0$ must be 
much smaller than the disorder broadening $\hbar/\tau$, and the other is the distance 
between two FSs $m\alpha/\hbar^2$ must be much larger than the shift of the 
distribution function in the momentum space $\tau e E /\hbar$ to avoid the level mixing by 
the applied electric field. The upper limit of the electric field is of the order of  $10^3$ V/m 
for GaAs, $10^{6\sim 7}$ V/m for BiTeI and $10^{5\sim 6}$ V/m for Bi$_2$Se$_3$; the 
latter two values are sufficiently large for usual terahertz experiments ($\sim10^5$ V/m). 
Note that two FSs come very close when $\alpha~\sim\beta$. In this case, the distance 
between two FSs in the momentum space becomes $m|\alpha-\beta|/\hbar^2 \sim 0$, 
and hence our results are not valid near the persistent helix phase, $\alpha=\beta$.

\textit{Acknowledgment ---}
We thank M. Kawasaki and Y. Tokura for fruitful discussions. 
This work was supported by the Grants-in-Aid for Scientic Research from MEXT KAKENHI (Grant Nos.JP25400317, JP15H05854 and JP17K05490) (ME), 
the Gordon and Betty Moore Foundation's EPiQS Initiative Theory Center Grant (TM),
and JSPS Grant-in-Aid for Scientic Research (No. 24224009, and No. 26103006) from MEXT, Japan,
and ImPACT Program of Council for Science, Technology and Innovation (Cabinet office, Government of
Japan) (NN).
KWK acknowledges support from “Overseas Research Program for Young Scientists” through Korea Institute for Advanced Study (KIAS).
This work is also supported by CREST, JST (JPMJCR16F1).

\appendix
\section{Derivation of second order spin current}
In this section, we show the derivation of the spin currents; $j_{x,s_{x}}^{(2)},j_{y,s_{x}}^{(2)},j_{y,s_{y}}^{(2)}$ for the surface of 3D TI and the Rashba system which are skipped in the main text.

\subsection{Surface of 3D TI}
For surface states of 3D TIs, the spin currents are given by
\begin{eqnarray}
j_{x,s_{x}}^{(2)} &=&\int \frac{\td^{2}\boldsymbol{k}}{{(2\pi )}^{2}}\ \left\langle \pm ,\boldsymbol{k} \left\vert \hat{\jmath}_{xs_{x}}\right\vert \pm ,\boldsymbol{k} \right\rangle f_{2}^{\pm }  \notag \\
&=&\int \frac{\td^{2}\boldsymbol{k}}{{(2\pi )}^{2}}\ 
\left[ \mp \frac{1}{2}\frac{k}{m}\sin \phi \cos \phi \right] \notag \\
& & \qquad\qquad \times\left[
e^{2}\tau ^{2}E^{2}\frac{\partial }{\partial k_{x}}\left( \frac{k}{m}\pm
v \right) \cos \phi \df\right] \notag \\
&=& \pm\frac{e^{2}\tau ^{2}E^{2}}{8\pi ^{2}}\int \td^{2}\boldsymbol{k}\ 
\frac{\sin ^{3}\phi }{m}\left[ \left( \frac{k}{m}\pm v \right) 
\cos\phi \df\right] \nonumber\\
&=&0,
\end{eqnarray}

\begin{eqnarray}
j_{y,s_{x}}^{(2)} &=&\int \frac{\td^{2}\boldsymbol{k}}{{(2\pi )}^{2}}\ 
\left\langle \pm ,\boldsymbol{k} \left\vert \hat{\jmath}_{ys_{x}}\right\vert
\pm ,\boldsymbol{k} \right\rangle f_{2}^{\pm }  \notag \\
&=& \int \frac{\td^{2}\boldsymbol{k}}{{(2\pi )}^{2}}\ 
\left[ \frac{1}{2}\left( \pm \frac{k}{m}\sin ^{2}\phi -v \right) \right]\notag\\
& &\qquad\qquad \times \left[ e^{2}\tau ^{2}E^{2}\frac{\partial }{\partial k_{x}}\left( \frac{k}{m}\pm v \right) 
\cos \phi \df \right]   \notag \\
&=&\mp \frac{e^{2}\tau ^{2}E^{2}}{8\pi ^{2}}\int \td^{2}\boldsymbol{k}\ 
\left[ \frac{1}{m}\cos \phi \sin ^{2}\phi \right] \notag\\
& &\qquad\qquad \times \left[ \left( \frac{k}{m}\pm v \right) \cos \phi 
\df\right]   \notag \\
&=&\mp \frac{e^{2}\tau ^{2}E^{2}}{32\pi m}\int k\td k\    \left( \frac{k}{m}\pm v \right) \df \notag \\
&=& \pm\frac{e^{2}\tau ^{2}E^{2}}{32\pi m} \left[ - mv + \sqrt{2m\mu+ m^2v^2} \right]  \notag \\
&=&\frac{1}{5}j_{x,s_{y}}^{(2)} 
\end{eqnarray}
and
\begin{eqnarray}
j_{y,s_{y}}^{(2)}&=&\int \frac{\td^{2}\boldsymbol{k}}{{(2\pi )}^{2}}\ 
\left\langle \pm ,\boldsymbol{k} \left\vert \hat{\jmath}_{ys_{y}}\right\vert
\pm ,\boldsymbol{k} \right\rangle f_{2}^{\pm }\notag\\
&=&-\int \frac{\td^{2}\boldsymbol{k}}{{(2\pi )}^{2}}\ 
\left\langle \pm ,\boldsymbol{k} \left\vert \hat{\jmath}_{xs_{x}}\right\vert \pm ,\boldsymbol{k} 
\right\rangle f_{2}^{\pm }\notag\nonumber \\
&=&-j_{x,s_{x}}^{(2)}=0.
\end{eqnarray}

\subsection{Rashba system}
For Rashba systems, the spin currents are given by
\begin{eqnarray}
j_{x,s_{x}}^{(2)R} &=&\sum_{\pm}\int \frac{\td^{2}\boldsymbol{k}}{{(2\pi )}^{2}}\ \left\langle \pm ,\boldsymbol{k} \left\vert \hat{\jmath}_{xs_{x}}\right\vert \pm ,\boldsymbol{k} \right\rangle f_{2}^{\pm }  \notag \\
&=&\sum_{\pm}\int \frac{\td^{2}\boldsymbol{k}}{{(2\pi )}^{2}}\ 
\left[ \mp \frac{1}{2}\frac{k}{m}\sin \phi \cos \phi \right]\notag\\
& &\qquad\qquad \times \left[
e^{2}\tau ^{2}E^{2}\frac{\partial }{\partial k_{x}}\left( \frac{k}{m}\pm
\alpha \right) \cos \phi \df\right] \notag \\
&=& \sum_{\pm}\frac{e^{2}\tau ^{2}E^{2}}{8\pi ^{2}}\int \td^{2}\boldsymbol{k}\ 
\frac{\sin ^{3}\phi }{m}\left[ \left( \frac{k}{m}\pm \alpha \right) 
\cos\phi \df\right] \notag\\
&=&0,
\end{eqnarray}

\begin{eqnarray}
j_{y,s_{x}}^{(2)R} &=&\sum_{\pm}\int \frac{\td^{2}\boldsymbol{k}}{{(2\pi )}^{2}}\ 
\left\langle \pm ,\boldsymbol{k} \left\vert \hat{\jmath}_{ys_{x}}\right\vert
\pm ,\boldsymbol{k} \right\rangle f_{2}^{\pm }  \notag \\
&=&\sum_{\pm} \int \frac{\td^{2}\boldsymbol{k}}{{(2\pi )}^{2}}\ 
\left[ \frac{1}{2}\left( \pm \frac{k}{m}\sin ^{2}\phi -\alpha \right) \right]\notag\\
& &\qquad\qquad \times
\left[ e^{2}\tau ^{2}E^{2}\frac{\partial }{\partial k_{x}}\left( \frac{k}{m}\pm \alpha \right) 
\cos \phi \df\right]   \notag \\
&=&\sum_{\pm}\mp \frac{e^{2}\tau ^{2}E^{2}}{8\pi ^{2}}\int \td^{2}\boldsymbol{k}\ 
\left[ \frac{1}{m}\cos \phi \sin ^{2}\phi \right]\notag\\
& &\qquad\qquad \times\left[ \left( \frac{k}{m}\pm \alpha \right) \cos \phi 
\df\right]   \notag \\
&=&\sum_{\pm}\mp \frac{e^{2}\tau ^{2}E^{2}}{32\pi m}\int k\td k\    \left( \frac{k}{m}\pm \alpha \right) \df  \notag \\
&=&- \frac{e^{2}\tau ^{2}E^{2}}{16\pi m}\times\left\{ 
\begin{array}{ll}
m\alpha & (\mu >0) \\ 
\sqrt{2m\mu+m^2\alpha^2}& (\mu <0)
\end{array}
\right.   \notag\\
&=& \frac{1}{5}j_{x,s_{y}}^{(2)R}
\end{eqnarray}
and
\begin{eqnarray}
j_{y,s_{y}}^{(2)R}&=&\int \frac{\td^{2}\boldsymbol{k}}{{(2\pi )}^{2}}\ 
\left\langle \pm ,\boldsymbol{k} \left\vert \hat{\jmath}_{ys_{y}}\right\vert
\pm ,\boldsymbol{k} \right\rangle f_{2}^{\pm } \notag\\
&=&-\int \frac{\td^{2}\boldsymbol{k}}{{(2\pi )}^{2}}\ 
\left\langle \pm ,\boldsymbol{k} \left\vert \hat{\jmath}_{xs_{x}}\right\vert \pm ,\boldsymbol{k} 
\right\rangle f_{2}^{\pm }\notag\\
&=&-j_{x,s_{x}}^{(2)R}=0.
\end{eqnarray}

\subsection{$1/m$ expansion in the surface of TI}
In this subsection, we investigate the effect of the parabola term in the Hamiltonian of the surface of 3D TI. 
When we expand the second order spin current with respect to $1/m$, we obtain
\begin{eqnarray}
j_{x,s_y}^{(2)}&=&5j_{y,s_x}^{(2)} \notag\\
&=& \pm\frac{5e^2\tau^2E^2}{32 \pi m} \left[ - mv + \sqrt{m^2v^2 +2m\mu}\right] \notag\\
&=& \pm\frac{5e^2\tau^2E^2v}{32 \pi}  \left[ \frac{\mu}{ mv^2}-\frac{1}{2} \left(\frac{\mu }{mv ^2}\right)^2 \right.\notag\\
& & \qquad \qquad \left.+\frac{1 }{2} \left(\frac{\mu }{mv ^2}\right)^3+O\left(\left(\frac{\mu}{mv^2}\right)^4\right) \right]\notag\\
&=& \pm\frac{5\mu e^2\tau^2E^2}{32 \pi mv}  + O\left(\left(\frac{\mu}{mv^2}\right)^2\right).
\end{eqnarray}
The leading term is the expression for $j^{(2)}_{x s_y}$, which is obtained only by considering the correction to the current operator due to the $k^2$ dispersion. Namely, when we evaluate the expectation value $\left\langle I,\bm{k}\left\vert \hat{\jmath}_{\mu s_{\nu}}\right\vert I,\bm{k} \right\rangle$,
we use  $\hat{\mathcal{H}}=k^2/(2m) + v(k_x\sigma_y-k_y\sigma_x)$ for the Hamiltonian but 
 $\varepsilon^\pm = \pm vk$ for the distribution function. In this situation, for example, $j_{x,s_y}^{(2)}$ is given by
\begin{eqnarray}
j_{x,s_y}^{(2)} &=&  \volintpi{\kk}{2} \ \left\langle \pm,\boldsymbol{k} \left\vert \hat{\jmath}_{\mu s_{\nu}}\right\vert \pm ,\boldsymbol{k} \right\rangle f_2 \notag\\
&=& \volintpi{\kk}{2} \ \left[\pm \frac{1}{2} \frac{k}{m} \cos ^2\phi +v\right]  \notag\\
& & \qquad\qquad\left[ e^{2}\tau ^{2}E^{2}\frac{\partial }{\partial k_{x}}\left(\pm v \right) \cos \phi \df \right]  \notag\\
&=&-\frac{v e^2\tau^2E^2}{8\pi m}\int \td^{2}\kk  \left[\cos^3\phi+2\sin^2\phi\cos\phi\right]\notag\\
& &\qquad\qquad \times\cos \phi \df \notag\\
&=&-\frac{5v e^2\tau^2E^2}{32\pi m}\int k\td k  \df \notag \\
&=&\frac{5v e^2\tau^2E^2}{32\pi m}\int k\td k  \delta (\pm v k-\mu )\notag \\
&=& \pm\frac{5\mu e^2\tau^2E^2}{32\pi mv} .
\end{eqnarray}
 This indicates that the spin current at the TI surface arises from the interplay between  the surface Weyl state exhibiting a nontrivial spin texture and the effect of the $k^2$ dispersion introducing the $k$ linear term in the current operator.

\subsection{Numerical calculation in Rashba-Dresselhaus system}
The second order spin current in the coexistence of the Rashba and the Dresselhaus terms, for example, $j_{x,s_x}^{(2)}$ is calculated as
\begin{eqnarray}
j_{x,s_x}^{(2)} &=&\sum_{\pm} \volintpi{\kk}{2} \ \left\langle \pm ,\boldsymbol{k} \left\vert \hat{\jmath}_{xs_{x}}\right\vert \pm ,\boldsymbol{k} \right\rangle  f^\pm_2 \notag\\
&=&\sum_{\pm} \volintpi{\kk}{2} \  \left[\frac{1}{2}  \left( \pm\frac{k}{m} \cos \phi\cos\varphi +\beta \right) \right]\notag\\
& &\qquad\qquad \times \left[e^2\tau ^2  \left(\dE \right)  \left( \dEene \right) \df\right] \notag \\
&=&\sum_{\pm} \pm \frac{e^2\tau^2}{8\pi^2 m}\int \ k \td k \td\phi \ \left( \dEene \right) \delta(\varepsilon^\pm-\mu)\notag\\
& &\qquad\qquad \times \left(\dE \right) \left[k \cos \phi\cos\varphi \right] .
\end{eqnarray}
The analytical integration over $k$ is possible for given values of $\phi$ with the use of the relations
\begin{equation}
\delta(\varepsilon^+-\mu) =\left\{ 
\begin{array}{ll}
\left| \frac{k}{m}+A \right| ^{-1}\delta\left( k-k_{F+}^+\right) & (\mu >0) \\ 
0 & (\mu <0 ) \\
\end{array}
\right.,  
\end{equation}
\begin{eqnarray}
& &\ \ \delta(\varepsilon^--\mu) = \notag \\
& &\left\{ 
\begin{array}{ll}
 \left| \frac{k}{m}-A \right| ^{-1}\delta\left( k-k_{F-}^+\right)& (\mu >0) \\ 
\left| \frac{k}{m}-A \right| ^{-1}\delta\left( k-k_{F-}^+\right) \\
 \ \ + \left| \frac{k}{m}-A \right| ^{-1}\delta\left( k-k_{F-}^-\right) & (\mu <0 , m^2A^2+2m\mu >0) \\
0 &  (\mu <0 , m^2A^2+2m\mu <0)  \\
\end{array}
\right. 
\end{eqnarray}
Then, the integral over $\phi$ is evaluated numerically. The similar calculations are carried out for the other components of the second order spin current.
This explassion is used in the Fig. 3 in the main text.

%\begin{thebibliography}{99}

%\end{thebibliography}

\end{document}